\documentclass[a4paper,12pt]{article}
\begin{document}
\begin{center}

{\bf A quest for weak objects and for invisible stars}

\vspace{0.4cm}

            S.I.Blinnikov$^{1,2,3}$

\vspace{0.2cm}

 $^1$ Institute for Theoretical and Experimental Physics\\
   117259, Moscow, Russia

\vspace{0.2cm}

$^2$ Sternberg Astronomical Institute, 119899 Moscow, Russia
\vspace{0.2cm}

$^3$ National Astronomical Observatory, Mitaka, Tokyo, Japan
\vspace{0.5cm}

     Presented at
 
                          Atami Workshop on

    BARYONIC MATTER IN THE UNIVERSE AND ITS SPECTROSCOPIC STUDIES

           (November 22 - 24, 1996, Atami, Japan)
\end{center}
\vspace{0.7cm}

High quality spectra
to be produced by SUBARU 8 meter telescope will be extremely valuable
for numerous astronomical projects. Here I concentrate on two classical
types of weak broad line objects -- distant supernovae
and white dwarfs -- with emphasis of their importance for fundamental
physics, not only for astronomy. With regard to recent results of
MACHO experiment and HST weak star counts  I discuss one exotic
possibility: invisible stars made of the so called ``mirror'' matter.

\section{Supernovae}

I will not attempt to review all problems of supernova theory
solution of which depends
heavily on the high quality spectra of those events. But one problem,
namely, a direct determination of the Hubble parameter, $H_0$,
with the help of supernova (SN) spectra, deserves special attention.
The value of $H_0$ is of paramount importance both for astronomy and
for physics.
The Hubble diagram for type Ia SNe is normally used to this end, when
they are treated as `standard candles',
as done e.g. by Sandage \& Tammann \cite{ST}.
Instead of this, a direct physical technique,
an `Expansion Photosphere Method' (EPM)
or its modifications, is used by a number of workers
to measure the distance to the host galaxy,
where SN occurs. SNe Ia are brighter (on average) than SNe II but
the physics of their light is much more complicated than that of type II
SNe: already near the maximum light SNe Ia do not have a true photosphere,
their continua being formed by blends of weak lines excited in
the conditions which deviate strongly from local thermodynamic
equilibrium. Type II SNe show a rich variety of light curves
and they clearly are not the `standard candles', but the presence of
hydrogen provides for a real photosphere for a couple of months in
many classical ``plateau'' light curve events. Even the peculiar
type II supernovae, SN 1987A
and SN 1993J, are modeled and understood much better than any of SNe Ia
in spite of many important achievements in the latter field, see e.g. refs.
\cite{Woo,Nom}.

It seems that EPM is not hard to implement.  Using weak lines which are
believed to be formed near the photosphere one can measure the photospheric
speed $v_{\rm ph}$ and then find its radius,
$$
 R_{\rm ph}=v_{\rm ph} t \; ,
$$
where $t$ is time measured after the appropriate zero moment. Now the distance
$D$ to the supernova is
$$
 D=R_{\rm ph}\sqrt{F_\lambda({\rm model}) \over F_\lambda({\rm observed})}
$$
if a reliable model flux $F_\lambda({\rm model})$ at the SN photosphere
is compared at the Earth's surface
with the flux $F_\lambda({\rm observed})$.

In spite of seeming simplicity,
the most recent determinations of $H_0$ by this method still give discordant
results:
B.Schmidt et al. \cite{Sch} find  $H_0=73\pm 6$, while
E.Baron et al. \cite{Bar} claim $H_0<50$. One of the reasons of this
discrepancy can be the prediction of higher $F_\lambda({\rm model})$ in
\cite{Bar} due to an underestimate of the expansion effect in line opacity
\cite{BHM} (see more accurate treatment of this effect in \cite{Blinn}).
In principle the spectrum fitting technique used in \cite{Bar} must give
more accurate results than the classical EPM which uses only continuum
fluxes, but two things are necessary for a successful implementation of
this technique: a full self-consistent hydrodynamic model of the supernova
light curve, producing a very good fit at least to a broad band photometry
(not a model power law atmosphere and not
only a good bolometric light curve fit),
and a very good low noise spectrum. While the photometry can be done
on a relatively modest telescope for a very distant supernova, a high
quality spectra (on the level of the spectra that theorists have for SN 1987A)
require a full power of the SUBARU telescope.

\section{White dwarfs}

Another group of weak, broad line objects are white dwarf (WD) stars.
The importance of old WDs is well known, and there may exist also
a population of `very ancient' WDs \cite{Alck}, but I wish to concentrate
on relatively young, age less $\sim 0.5$ Gyr, and hot $T_{\rm eff} > 
12 \times 10^3$ K, white dwarfs.
A search of hot WDs \cite{FLG,LDM} is independent
of the proper motions and can be much deeper than proper motion surveys
of cold WDs. It can be combined with a survey of quasars like in
the Homogeneous Bright QSO Survey (HBQS) project  \cite{Gemmo}.

Good spectra of very distant WDs (of DA and DO classes)
together with model atmospheres give
very valuable information. One can obtain then
a reliable estimate of bolometric magnitude $M_{\rm bol}$,
of effective temperature $T_{\rm eff}$ and gravity $g_{\rm eff}$. This
opens a way to estimate the gravitational redshift (because mass and
radius are reliably estimated from theory, given $g_{\rm eff}$), and
to separate it from the Doppler shift. Wide, pressure broadened, lines
of WDs hinder redshift measurements, but with a good resolution one
can see very narrow cores of WD lines (due to deviations from LTE) and to
see all those effects (as well as WD rotation).

With the power of SUBARU telescope a survey of hot WDs can be carried
out to a very high galactic $z$ coordinate and bring data on their
distribution, kinematics, birthrate, progenitors etc. Note, that
among the progenitors of hot WDs there must be not only ``contemporary''
main sequence stars, but also the oldest stars of our Galaxy which
are completing their evolution just now. To find this subpopulation
among hot WDs one needs a rich statistics at high $z$ and data on kinematics
since these stars are expected to be faster than those stemming from
younger predecessors. With accurate data on $g_{\rm eff}$ one can find
WD distribution over mass. The median value of the mass distribution
can be checked by another independent method \cite{BDB}. The evolution
of hottest WDs is controlled by neutrino emission: higher is the mass,
higher is the density and the neutrino luminosity. Thus, more massive WDs
spend less time at high $T_{\rm eff}$,  and the $T_{\rm eff}$  distribution
of a population of hot WDs can be used to derive the median mass
of the population.

The $T_{\rm eff}$  distribution is extremely valuable for constraining
electromagnetic properties of neutrinos, or new exotic particles,
such as axions (see \cite{BDB} and references therein). Broad band
photometry used in surveys \cite{FLG,LDM} is not adequate to give
the richness of information contained in WD spectra. One can hope
to reach the goals outlined here only expanding surveys based on
spectra as started in HBQS survey \cite{Gemmo}. Only
a fraction of the material on WDs is published in the latter reference.
It is hard to believe, but this project is suspended (or stopped
altogether), and a major part of the spectral material is not yet
processed. In my opinion, this work deserves attention and support
of SUBARU team, and the experience acquired by
A.Gemmo and her colleagues \cite{Gemmo} should be employed in
future collaboration projects.

\section{Invisible stars}

Recent MACHO project microlensing results \cite{Alck,AMS}  imply an
appreciable fraction of dark matter in the form of compact objects with
the typical mass $0.5^{+0.3}_{-0.2}M_\odot$. It is suggested \cite{AMS}
that they can be a population of very ancient white dwarfs. The number
of those WDs must be huge (they can contain  between $\sim 20$\%
and 100\% of the halo dark matter \cite{AMS}) and if they  really exist this
will involve many problems in explaining their initial mass function,
chemical evolution of the Galaxy, early supernovae etc. So other, more
exotic explanations of MACHO results, like primordial black holes
or ``neutralino'' stars, are possible. To me, it seems most natural
to explain the microlensing results by the presence of the population
of invisible stars made of the so called ``mirror'' matter.

The concept of the mirror particles stems from the idea of  
Lee and Yang \cite{LY} who suggested the existence of new particles
with the reversed sign of the mirror asymmetry observed in
our world. Lee and Yang \cite{LY} believed that the new particles
(whose masses are degenerate with the masses of ordinary particles)
could participate in the ordinary interactions. Later, 
Kobzarev, Okun \& Pomeranchuk \cite{KOP} have shown that ordinary strong,
weak and electromagnetic interactions are forbidden for the mirror particles 
by experimental evidence, only gravity and super-weak interaction
is allowed for their coupling to the ordinary matter. But if they
really mirror the properties of ordinary particles, this means that there
must exist mirror photons, gluons etc., coupling the mirror fermions
to each other, like in our world. Thus the possibility of existence
of the mirror world was postulated first in the ref. \cite{KOP}.
The particle mass pattern and particle interactions in the mirror world
are quite analogous to that in our world, but the two worlds interact
with each other essentially through gravity only.

Later the idea was developed in a number of papers, e.g.
\cite{Okun,BlKh},
and the interest to it is revived recently in attempts to explain all
puzzles of neutrino observations \cite{Foot,Ber}. It is shown in the cited
papers that a world of mirror particles can coexist with our, visible, world,
and some effects that should be observed are discussed.

It is shown in ref. 
\cite{BlKh} that ordinary and mirror matter are most likely well mixed on the 
scale of galaxies, but not in stars, because of
different thermal or gasdynamic processes like SN shock waves which induce
star formation. It is predicted in \cite{BlKh} that star counts by HST
must reveal the deficit of local luminous matter if the mirror stars
do really exist in numbers comparable to ordinary stars and contribute
to the gravitational potential of galactic disk. 
New HST results \cite{GBF} show the reality of the luminous
matter deficit: e.g., instead of 500 stars expected from Salpeter mass
function in the HST fields investigated for the range of absolute
visual magnitudes $ 14.5 < M_V < 18.5 $
only 25 are actually detected. It is found that the Salpeter slope
does not continue down to the hydrogen-burning limit but has a maximum
near $M \sim 0.6 M_\odot$, so lower mass stars do not contribute
much to the total luminous mass as was thought previously.
The total column density of the galactic disk, $\Sigma \approx 40
M_\odot {\rm pc}^{-2}$ is a factor of two lower than
published estimates of the dynamical mass of the disk \cite{GBF}.
It should be
remembered that here we discuss a contribution of invisible stars to
the gravity of the galactic disk which has more to do with the local Oort
limit \cite{Oort} than with the halo dark matter.

Let us turn now to the implications of microlensing results for the searches
of mirror matter. I believe that one historical remark is in order. The 
classical reference here is the paper  \cite{Rfd}, but Refsdal
suggested to follow a lensing star as the line of sight approaches
a distant source to detect the effect \cite{Rfd}. Perhaps the first
work which discussed the microlensing on invisible stars is the paper
by Byalko \cite{Byal}. He has derived the effect accurately and pointed
out the importance of observing the bulge region of our Galaxy in order
to derive ``information on the nature of scattering (i.e. lensing) 
objects'' \cite{Byal}.

What can one predict for MACHOs if the mirror world is real? If the 
properties of the mirror particles are {\em exact} mirror reflections of
our particles, then it is unlikely that 100\% of all halo dark matter
consists of the mirror stars. It is more natural to expect that in this
case the latter are distributed like visible stars in our Galaxy with
their spheroidal and disk populations\cite{BlKh}, and the bulk of the halo dark matter
consists of massive neutrinos. The  development of the idea of
the mirror world \cite{Okun} (see recent papers \cite{Ber})
suggest that the symmetry is not so
strict, but again it is natural to expect that halo consists mostly
of massive neutrinos. So in the scenarios discussed above one should
expect that MACHOs from mirror stars are more concentrated to the
galactic disk and to the bulge than distributed more uniformly in the
halo. Of course, the exact properties of mirror particles and the
evolution of the mirror world are not known to anybody and one can picture
to oneself our spiral Galaxy embedded into a giant elliptical consisting 
of mirror stars etc. What is needed now, is not a speculation, but a richer
statistics of MACHO events, with a goal to derive mass and space
distributions of lensing objects, and the  search of very weak stars on the
most powerful telescopes (the effects of the accretion of ordinary,
`visible' plasma onto mirror WDs produce a very low luminosity
$\sim 10^{-6} L_\odot \rho_{-24}/v^3_6$ with the spectrum which is
not yet calculated by anybody, to the best of my knowledge).

In summary, if the search for very
ancient degenerate dwarf stars by SUBARU telescope, suggested by C.Alcock
\cite{Alck}
at this meeting, will give a negative result (or a very low number of those
stars), then the possibility of real existence of mirror stars will
not seem so exotic.

{\bf Acknowledgements.} I am very grateful to NAO for support of my
visit to Japan and I cordially thank T.Kajino and K.Nomoto for their
hospitality. I gratefully acknowledge stimulating discussions with L.B.Okun
and I thank him for improving the presentation. 
My work in Russia is partly supported by RBRF grants 96-02-16352 and
96-02-19756.

\end{document}